\documentclass[sigconf]{acmart}
\usepackage{multirow}
\AtBeginDocument{%
  }
\copyrightyear{2026}
\acmYear{2026}
\setcopyright{cc}
\setcctype{by}
\acmConference[E-Energy '26]{The 17th ACM International Conference on Future and Sustainable Energy Systems}{June 22--25, 2026}{Banff, AB, Canada}
\acmBooktitle{The 17th ACM International Conference on Future and Sustainable Energy Systems (E-Energy '26), June 22--25, 2026, Banff, AB, Canada}
\acmDOI{10.1145/3744255.3811735}
\acmISBN{979-8-4007-2011-6/2026/06}

\begin{document}

\title[GPU-Native Multi-Area State Estimation]{GPU-Native Multi-Area State Estimation via SIMD Abstraction and Boundary Condensation}

\author{Yifei Xu}
\affiliation{%
  \institution{New York University}
  \city{New York}
  \country{USA}
}
\email{yifeixu@nyu.edu}

\author{Yuzhang Lin}
\affiliation{%
  \institution{New York University}
  \city{New York}
  \country{USA}
}
\email{yuzhang.lin@nyu.edu}
\renewcommand{\shortauthors}{Xu and Lin}

\begin{abstract}
Power system state estimation (SE) is foundational for grid monitoring, yet conventional centralized solvers face increasing computational pressure as the system scale and real-time requirements grow. This paper presents a GPU-native framework for hierarchical multi-area state estimation (MASE) that addresses these bottlenecks through a single-instruction, multiple-data (SIMD) abstraction and sparse Schur local condensation. We partition the network into areas, evaluate measurement residuals and derivatives using fixed-sparsity templates, and directly assemble local normal-equation blocks through a fused GPU accumulation kernel without materializing explicit Jacobians. Each area is then factorized on the GPU in Schur mode to export a dense local boundary block and condensed right-hand side, after which a reduced global boundary system is assembled and solved on device. This design preserves device residency across measurement evaluation, local condensation, and boundary coordination while exposing parallelism across areas. Numerical experiments on partitioned PEGASE 2869-bus, PEGASE 9241-bus, and ACTIVSg10k benchmark systems demonstrate that the proposed approach effectively leverages GPU throughput by maintaining full device residency and high arithmetic intensity.
\end{abstract}




\maketitle

\section{Introduction}

Power system state estimation (SE) produces a statistically consistent snapshot of system states from redundant, noisy measurements and is a cornerstone of energy management systems \cite{abur2004power}. As transmission networks expand and measurement granularity increases, the growing computational burden of centralized SE poses a significant challenge to operational responsiveness and downstream applications \cite{gomez2011taxonomy, zhao2005multi, yilmaz2023robust}.

A well-established resolution to scaling pressure is multi-area state estimation (MASE), where the network is partitioned into areas and local estimators cooperate through a decomposition--coordination mechanism \cite{gomez2011taxonomy, guo2016hierarchical, wang2020privacy, zhao2005multi}. A critical review classifies MASE methods into hierarchical versus decentralized approaches and highlights scalability and data locality as primary motivations \cite{gomez2011taxonomy}. While these frameworks distribute the problem size, the underlying computational engine often remains unchanged. Existing MASE implementations typically rely on legacy sparse linear algebra kernels, where every Gauss–Newton (GN) iteration incurs significant overhead from Jacobian re-assembly and sparse factorization. Consequently, the architectural benefits of partitioning are diluted by the latency of these serial, memory-bound operations.

While recent GPU-native frameworks \cite{shin2024accelerating, lu2025cupdlp} have proven that expressing models as single-instruction, multiple-data (SIMD) templates significantly reduces per-iteration overhead, these insights have not yet been incorporated to MASE. On one hand, the MASE literature predominantly focuses on decomposition strategies while relying on legacy CPU-centric sparse pipelines. On the other hand, generic GPU tools do not address the specific challenges of SE, such as the repeated linearization of graph-structured measurements or the requirement for end-to-end device residency during boundary coordination. This paper therefore co-designs the measurement abstraction, local elimination backend, and coordinator assembly for the multi-area SE pipeline.

The contributions of this paper are summarized as follows:
\begin{itemize}
\item We formulate a boundary-condensed hierarchical MASE algorithm that eliminates local internal variables via Schur complements, yielding a compact reduced boundary system for efficient global coordination.
\item We design a GPU-resident execution pipeline that combines fixed-sparsity SIMD measurement templates with fused accumulation of local normal equations, avoiding explicit Jacobian materialization and minimizing host--device traffic. The local sparse elimination is further realized with a GPU Schur-mode backend.
\end{itemize}

\section{Background and Problem Setup}
\label{sec:background}

\subsection{WLS SE and GN Updates}
We consider the standard nonlinear measurement model
\begin{equation}
z = h(x) + e,
\label{eq:meas_model}
\end{equation}
where $x\in\mathbb{R}^{n}$ denotes the system state, $z\in\mathbb{R}^{m}$ denotes the measurement vector ($m>n$), $h(\cdot)$ denotes the nonlinear measurement functions, and $e$ denotes the measurement error vector, which is commonly modeled as zero-mean Gaussian with covariance $R$ \cite{abur2004power}.

In the conventional bus-branch formulation, $x$ stacks bus voltage magnitudes and phase angles, while $z$ includes voltage magnitudes, power injections, and branch power flows; phasor measurement unit (PMU) measurements can be incorporated as additional measurement functions.

The weighted least-squares (WLS) estimator minimizes the objective function defined in matrix form as
\begin{equation}
  \min_{x} \; J(x) = r(x)^{\top} W r(x), \qquad r(x) = z - h(x),
  \label{eq:wls}
\end{equation}
where $W$ is the measurement weight matrix (typically $W=R^{-1}$).

A GN iteration updates $x_{k+1}=x_k+\Delta x_k$ by solving the normal equations
\begin{equation}
  G_k \Delta x_k = H_k^{\top} W r(x_k), \quad
  H_k = \left.\frac{\partial h}{\partial x}\right|_{x_k}, \quad
  G_k = H_k^{\top} W H_k,
  \label{eq:gn_normal}
\end{equation}
where $G_k$ is the gain matrix and $H_k$ is the Jacobian matrix.

\subsection{SIMD Measurement Abstraction}
To expose fine-grained parallelism suitable for GPU execution, we adopt the SIMD abstraction framework proposed in \cite{shin2024accelerating}. This framework classifies model evaluations into three computational patterns amenable to massive parallelism: 
\begin{itemize}
    \item Pattern 1 (Map), where a function is applied independently over data arrays;
    \item Pattern 2 (Reduction), where results are aggregated via a commutative operator (e.g., summation); 
    \item Pattern 3 (Scan), which handles recursive compositions.
\end{itemize}

We restructure the SE measurement equations to align strictly with these patterns. Let $\tau(i) \in \mathcal{T}$ denote the measurement type for index $i$ (e.g., voltage magnitude, active/reactive injection, branch flow), and let $p_i$ encompass the discrete parameters required for evaluation (e.g., bus indices, branch admittance). The measurement model is expressed as:
\begin{equation}
z_i = h_{\tau(i)}(x; p_i) + e_i, \quad i=1,\dots,m.
\label{eq:typed_meas}
\end{equation}
This formulation allows residual evaluation to be implemented as a parallel map operation, while the accumulation of Jacobian terms and GN right-hand sides follows a map-reduce pattern. It confers three distinct advantages: (i) the entire measurement set is processed by a minimal number of GPU kernels (one per type $\tau \in \mathcal{T}$); (ii) it facilitates type-specific code generation for analytic derivatives; and (iii) the symbolic sparsity pattern of the Jacobian remains invariant across iterations.
\subsection{Multi-area Decomposition and Boundary Variables}
We partition the network into $K$ areas interconnected by tie-lines. Boundary buses are the terminal buses of cut branches. For area $k$, state variables are classified as internal variables $x_i^{k}$ and boundary variables $x_b^{k}$. Local boundary vectors may overlap across areas, but they are linked to a single global boundary vector $x_{\Gamma}$ through local selector matrices. We focus on a hierarchical architecture where local processors estimate area states and a coordinator reconciles boundary coupling \cite{gomez2011taxonomy}. This setup is used in Section~\ref{sec:method} to derive the boundary-condensed coordination step.

\section{Boundary-Condensed MASE Method}
\label{sec:method}

This section details a hierarchical MASE algorithm based on Schur complement condensation. The central coordinator solves a reduced system involving only boundary variables, while each area performs local computations on its own measurements and states.

\subsection{Local Subproblems}
Let $x_{\Gamma}\in\mathbb{R}^{n_\Gamma}$ denote the global boundary state stacking voltage angles and magnitudes at boundary buses. For area $k$, boundary variables are selected by a (0,1) matrix $E_k$:
\begin{equation}
x_b^k = E_k x_{\Gamma},\qquad \Delta x_b^k = E_k \Delta x_{\Gamma}.
\label{eq:Ek_select}
\end{equation}
At coordinator iteration $t$, the coordinator broadcasts $x_{\Gamma}^{(t)}$ and each area solves a local WLS problem with fixed boundary:
\begin{equation}
x_i^{k,(t)} \in \arg\min_{x_i^k}\;
J_k\!\left(x_i^k;\,x_b^{k,(t)}\right),
\qquad x_b^{k,(t)} = E_k x_\Gamma^{(t)}.
\label{eq:local_wls}
\end{equation}
In practice, \eqref{eq:local_wls} is solved by a small number of GN steps in $x_i^k$ with $x_b^k$ treated as a parameter.

Let $r_k=z_k-h_k(x_i^k,x_b^k)$ denote the local residual and $W_k$ the local weight matrix. The Jacobian partition is
\begin{equation}
H_k = \left[\;H_i^k \;\; H_b^k\;\right],
\end{equation}
where $H_i^k=\partial h_k/\partial x_i^k$ and $H_b^k=\partial h_k/\partial x_b^k$.
The GN normal equations have the block form
\begin{equation}
\begin{bmatrix}
G_{ii}^k & G_{ib}^k\\
G_{bi}^k & G_{bb}^k
\end{bmatrix}
\begin{bmatrix}
\Delta x_i^k\\
\Delta x_b^k
\end{bmatrix}
=
\begin{bmatrix}
b_i^k\\
b_b^k
\end{bmatrix},
\label{eq:block_normal}
\end{equation}
with $G_{ii}^k=(H_i^k)^\top W_k H_i^k$, $G_{ib}^k=(H_i^k)^\top W_k H_b^k$, $G_{bb}^k=(H_b^k)^\top W_k H_b^k$, and $b_k=H_k^\top W_k r_k$.

\subsection{Boundary Condensation and Global Coordinator}
To decouple the internal variables, we eliminate $\Delta x_i^k$ using the first block row of \eqref{eq:block_normal}:
\begin{equation}
\Delta x_i^k = (G_{ii}^k)^{-1} \left( b_i^k - G_{ib}^k \Delta x_b^k \right).
\label{eq:backsub}
\end{equation}
Substituting into the second row gives the condensed boundary-only system:
\begin{equation}
S_b^k \, \Delta x_b^k = \hat{b}_b^k,
\label{eq:local_schur_system}
\end{equation}
where the Schur complement matrix $S_b^k$ and the modified right-hand side $\hat{b}_b^k$ are defined as:
\begin{equation}
S_b^k := G_{bb}^k - G_{bi}^k (G_{ii}^k)^{-1} G_{ib}^k, \qquad
\hat{b}_b^k := b_b^k - G_{bi}^k (G_{ii}^k)^{-1} b_i^k.
\label{eq:schur_defs}
\end{equation}

The coordinator then aggregates the local condensed systems into a global boundary model:
\begin{equation}
S_{\Gamma} := \sum_{k=1}^{K} E_k^\top S_b^k E_k, \qquad
\hat{b}_{\Gamma} := \sum_{k=1}^{K} E_k^\top \hat{b}_b^k.
\label{eq:global_assembly}
\end{equation}
Solving the reduced system $S_{\Gamma} \Delta x_{\Gamma} = \hat{b}_{\Gamma}$ yields the global boundary update and each area obtains $\Delta x_b^k = E_k\Delta x_\Gamma$. The internal update is then recovered from \eqref{eq:backsub}.

Although the derivation above is written for WLS/GN, the proposed structure is more general than plain WLS. Any estimator or system model whose local iteration yields a block-structured linear system in internal and boundary variables can use the same Schur-condensation idea. By contrast, estimators that lead to fundamentally different or indefinite KKT systems require modified local algebra and are outside the scope of the present implementation.

\section{GPU-Resident Implementation}
\label{sec:gpu}

The proposed framework achieves full device residency by mapping the partitioned MASE workflow onto the GPU. Parallelism is expressed at two levels: data-parallel template evaluation within each area and concurrent execution across areas during local Schur condensation.

\subsection{Fixed-Sparsity Measurement Templates}
\label{sec:templates}
To maximize arithmetic intensity and minimize memory latency, we perform a one-time symbolic dependency analysis to precompute the sparsity patterns of all measurement rows. Measurements are categorized into a minimal set of computational templates. For each template, we construct static index maps (e.g., neighbor lists derived from the local admittance matrix) and store measurement data in Structure-of-Arrays format. This abstraction decouples the sparsity structure from the numerical values, ensuring that memory access patterns remain invariant throughout the iterative GN process. 

Structured missing measurements are handled by preserving the fixed template slots and refreshing per-row weights so that masked rows contribute zero without rebuilding the template structure.

\subsection{Fused Accumulation of Normal Equations}
\label{sec:fused}
Standard sparse matrix assembly often incurs substantial overhead due to irregular memory writes and intermediate storage of the Jacobian matrix $H_k$. We circumvent these bottlenecks via a fused accumulation strategy. Dedicated GPU kernels iterate over the measurement templates to simultaneously evaluate residuals and analytic derivatives. Rather than materializing $H_k$, these kernels utilize atomic scatter-add operations to accumulate weighted contributions directly into the local normal equation blocks $G^k_{ii}, G^k_{iB}, G^k_{BB}$ and $b^k_i, b^k_B$. This approach effectively functions as a specialized map-reduce operation, maximizing cache locality and significantly reducing global memory traffic compared to traditional sparse triplet assembly.

\subsection{Sparse Local Schur Extraction}
\label{sec:schur_backend}
Each area maintains a persistent sparse Schur-mode factorization cache for the full local matrix \eqref{eq:block_normal}. The CSR sparsity pattern is constructed once, after which each GN iteration performs value-only updates, sparse numeric refactorization, dense export of the local Schur block, partial forward solve for condensed right-hand-side formation, and partial backward solve for interior recovery. Area-level condensation is issued on independent non-blocking CUDA streams \cite{cuda_toolkit}, which allows the sparse factorization and partial-solve phases of different areas to overlap.

\subsection{Fused Boundary Assembly and Global Solve}
\label{sec:boundary}
The coordination step is handled by a specialized fused assembly kernel that operates entirely on-device. This kernel maps local boundary indices to the global ordering using precomputed mapping arrays $E_k$, utilizing atomic operations to aggregate the global boundary system $S_\Gamma$ and $\hat{b}_\Gamma$. The reduced system is solved using a Cholesky factorization on the GPU. The boundary update is then scattered back to area-local buffers, after which the local sparse Schur caches perform the partial backward solves needed for interior recovery.


\section{Numerical Results}
\label{sec:eval}

\subsection{Experimental Setup}

We evaluate the proposed GPU-based MASE method on
three transmission-network benchmarks: PEGASE-2869 \cite{fliscounakis2013contingency}, PEGASE-9241 \cite{josz2016ac}, and ACTIVSg10k \cite{birchfield2016grid}. The solver is implemented in Python with CuPy for device memory management and custom CUDA kernels for measurement evaluation and fused accumulation; the implementation is available in an online repository \cite{multi_area_se_repo}. Local elimination is carried out in cuDSS Schur mode \cite{cudss2024}. All experiments were performed on a workstation with an AMD Ryzen 9 7950X3D CPU (16C/32T), an NVIDIA GeForce RTX 4080 GPU (16~GB), Linux 5.15, cuDSS 0.7.1.6, and CuPy 13.6.0. All reported results use FP64 precision.

Synthetic measurements include voltage magnitudes, active and reactive power injections, and branch power flows. The measurements are generated by adding i.i.d.\ Gaussian noise to an AC power-flow ground truth, with standard deviations of $0.01$ p.u.\ for voltage magnitudes and $0.02$ p.u.\ for power quantities. The solver uses Gauss--Newton updates with a maximum of 10 iterations. For each case, we report mean end-to-end solve times over 11 sequential runs and exclude the first run in order to separate one-time CUDA context creation, kernel compilation, and library initialization from steady-state throughput. All template indices, mapping arrays, and working buffers are allocated once and remain device-resident throughout the solve. 

We compare the proposed multi-area GPU solver against four baselines:
\begin{enumerate}
    \item \textbf{Centralized CPU (1T):} full WLS-GN with a sparse CPU solve and BLAS/OpenMP pinned to one thread;
    \item \textbf{Centralized CPU (6T):} the same centralized solver with explicit 6-thread BLAS/OpenMP settings;
    \item \textbf{Centralized GPU:} full WLS-GN with a monolithic cuDSS sparse GPU solve;
    \item \textbf{Multi-area CPU:} the same boundary-condensed algorithm executed with six worker processes and one BLAS/OpenMP thread per worker.
\end{enumerate}
The proposed method is reported as \textbf{Multi-area GPU}. Unless otherwise noted, the main comparisons use $K=6$ tree partitions.

\subsection{Benchmark Results}

\begin{figure}[ht]
\centering
\includegraphics[width=\columnwidth]{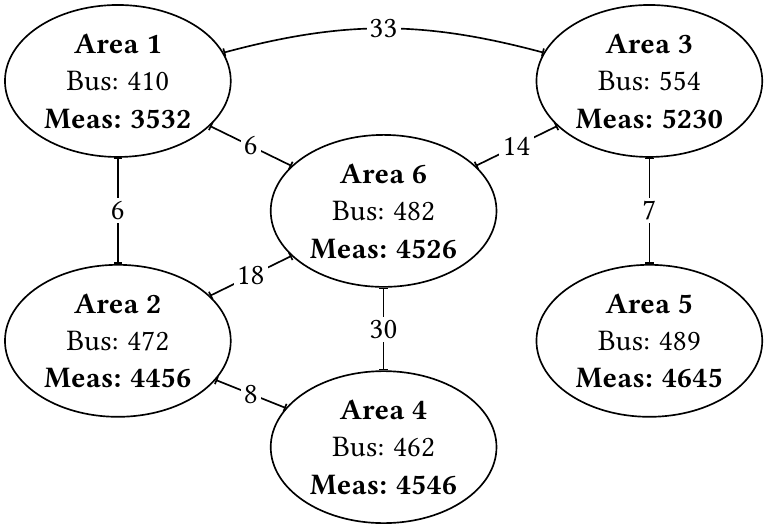}
\Description{Area-level graph for the six-area PEGASE-2869 partition. Six elliptical nodes labeled Area 1 through Area 6 show the number of buses and measurements in each area. Thick labeled edges between nodes indicate the number of tie-lines between adjacent areas.}
\caption{Area-level graph of the PEGASE-2869 partition. Node labels indicate the number of buses and local measurements in each area, and edge labels indicate the number of tie-lines between adjacent areas.}
\label{fig:case2869_area_graph_k6}
\end{figure}

We first describe the partition structure using PEGASE-2869, a large-scale pan-European transmission benchmark. The system contains 2869 buses and 4582 branches, yielding 5737 state variables in the bus--branch SE formulation. The network is partitioned into six connected areas with bus counts $(410,472,554,462,489,482)$, as shown in Fig.~\ref{fig:case2869_area_graph_k6}. This partition induces 122 cut branches and 172 boundary buses across 8 adjacent area pairs, yielding 343 boundary variables.

Table~\ref{tab:main_bench} summarizes the steady-state timing results on the three benchmark systems. The proposed multi-area GPU method achieves the lowest runtime in all three cases. On PEGASE-2869, the centralized CPU (1T), centralized CPU (6T), centralized GPU, and multi-area CPU baselines require 85.7~ms, 90.3~ms, 36.9~ms, and 82.8~ms, respectively,
whereas the proposed method attains 24.7~ms. This corresponds to a $3.47\times$ speedup over centralized CPU (1T), a $3.35\times$ speedup over multi-area CPU, and a $1.49\times$ speedup over centralized GPU.

The same pattern persists on the larger systems. On PEGASE-9241 and ACTIVSg10k,
the proposed solver attains 49.2~ms and 42.5~ms, respectively, corresponding to speedups of $6.90\times$ and $10.92\times$ over centralized CPU (1T), $5.85\times$ and $5.92\times$ over multi-area CPU, and $2.32\times$ and $2.89\times$ over centralized GPU. Although ACTIVSg10k is larger overall, PEGASE-9241 produces a less favorable partition geometry at $K=6$, with a larger boundary system and stronger inter-area coupling, which makes the condensed solve slightly more expensive.
Across all three cases, the centralized GPU baseline is already faster than the multi-area CPU baseline, which indicates that decomposition alone does not guarantee the best performance. Rather, the largest gains are obtained when the multi-area formulation is combined with GPU-resident local elimination and boundary coordination.

\begin{table*}[ht]
  \centering
  \small
  \caption{Steady-state end-to-end solve times on the three benchmark systems. Times are reported in milliseconds. Speedups are computed relative to the centralized CPU (1T) baseline.}
  \label{tab:main_bench}
  \begin{tabular}{llccc}
    \toprule
    Case & Method & Typical iters & Avg.\ excl.\ first (ms) & Speedup vs.\ 1T CPU \\
    \midrule
    \multirow{5}{*}{PEGASE-2869}
    & Centralized CPU (1T)              & 3--5 & 85.7 & 1.00$\times$ \\
    & Centralized CPU (6T)              & 3--5 & 90.3 & 0.95$\times$ \\
    & Centralized GPU                   & 3--5 & 36.9 & 2.32$\times$ \\
    & Multi-area CPU                    & 2--4 & 82.8 & 1.03$\times$ \\
    & Multi-area GPU      & 2--4 & 24.7 & 3.47$\times$ \\
    \midrule
    \multirow{5}{*}{PEGASE-9241}
    & Centralized CPU (1T)              & 3--4 & 339.5 & 1.00$\times$ \\
    & Centralized CPU (6T)              & 3--4 & 336.6 & 1.01$\times$ \\
    & Centralized GPU                   & 3--4 & 114.1 & 2.98$\times$ \\
    & Multi-area CPU                    & 3--4 & 287.8 & 1.18$\times$ \\
    & Multi-area GPU      & 3--4 & 49.2  & 6.90$\times$ \\
    \midrule
    \multirow{5}{*}{ACTIVSg10k}
    & Centralized CPU (1T)              & 4--5 & 464.2 & 1.00$\times$ \\
    & Centralized CPU (6T)              & 4--5 & 482.8 & 0.96$\times$ \\
    & Centralized GPU                   & 4--5 & 122.9 & 3.78$\times$ \\
    & Multi-area CPU                    & 3--4 & 251.6 & 1.84$\times$ \\
    & Multi-area GPU      & 3--4 & 42.5  & 10.92$\times$ \\
    \bottomrule
  \end{tabular}
\end{table*}

Across all three systems, the compared methods converge to the same solution within the reported precision. For PEGASE-2869, all five rerun variants produce essentially identical results, with a final objective of approximately 10,765.21 and a weighted residual norm of approximately 146.73. The same
consistency is observed on the larger systems: all methods on PEGASE-9241 converge to a final objective of approximately 37,078.30 with weighted residual norm 272.32, whereas all methods on ACTIVSg10k converge to a final objective of approximately 30,553.67 with weighted residual norm 247.20. These results confirm that the reported speedups reflect improved computational efficiency rather than changes in the converged estimate.

\subsection{Partition-Count Sweep}

To assess the sensitivity of the method to the number of partitions, Table~\ref{tab:ksweep} reports a partition-count sweep on PEGASE-9241. The results show a clear moderate-$K$ sweet spot rather than monotonic improvement with increasing partition count. For the multi-area GPU solver, the best runtime is obtained at $K=4$, with $K=3$ close behind. At $K=2$, the system remains under-partitioned and the local subproblems are still relatively large. In contrast, for $K \ge 6$, the boundary dimension grows substantially and an increasing fraction of the runtime is shifted to the coordinator. The multi-area CPU solver exhibits the same qualitative trend.

\begin{table}[ht]
  \centering
  \scriptsize
  \caption{Partition-count sweep on PEGASE-9241 using tree partitions.
  Coordinator share is defined as
  $(\text{boundary assembly}+\text{boundary solve})/\text{total}$.}
  \label{tab:ksweep}
  \begin{tabular}{cccccc}
    \toprule
    $K$ & Boundary dim & CPU total (ms) & CPU coord.\ share & GPU total (ms) & GPU coord.\ share \\
    \midrule
    2  & 128  & 504.9 & 0.2\%  & 56.4 & 2.4\% \\
    3  & 244  & 386.5 & 0.7\%  & 49.6 & 4.3\% \\
    4  & 366  & 355.8 & 1.7\%  & 44.6 & 6.8\% \\
    6  & 658  & 288.2 & 7.0\%  & 56.9 & 13.5\% \\
    8  & 766  & 367.0 & 7.9\%  & 61.9 & 13.1\% \\
    12 & 1602 & 495.1 & 36.4\% & 87.9 & 27.3\% \\
    \bottomrule
  \end{tabular}
\end{table}

The coordinator is therefore not the dominant cost in the best-performing regime. It becomes a first-order cost only in the over-partitioned regime, where the boundary dimension increases from 128 or 366 to 1602.

\subsection{Structured Masking Experiment}

To examine how the fixed measurement templates behave under missing measurements, we performed a structured masking experiment on PEGASE-2869 with $K=6$. In each experiment, one complete from/to-end flow-measurement family ($P_f$, $P_t$, $Q_f$, or $Q_t$) is removed. Each such masking pattern removes 4582 rows, corresponding to approximately 17.01\% of the total measurements. Table~\ref{tab:masking} shows that all masked variants remain numerically stable and exhibit essentially unchanged steady-state runtime. These results indicate that coherent block-wise measurement removal can be handled through row masking without rebuilding the template.

\begin{table}[ht]
  \centering
  \small
  \caption{Structured block masking on PEGASE-2869 with $K=6$ multi-area GPU solution. All cases converged in 5 iterations.}
  \label{tab:masking}
  \begin{tabular}{lcc}
    \toprule
    Masked family & Removed rows & Avg.\ excl.\ first (ms)\\
    \midrule
    None   & 0    & 26.2 \\
    $P_f$  & 4582 & 26.4 \\
    $P_t$  & 4582 & 25.6 \\
    $Q_f$  & 4582 & 28.2 \\
    $Q_t$  & 4582 & 26.6 \\
    \bottomrule
  \end{tabular}
\end{table}



\section{Conclusion}
\label{sec:conclusion}

In this note, we presented a GPU-native hierarchical MASE framework that combines fixed-sparsity SIMD measurement templates, fused accumulation of local normal-equation blocks, Schur-based local elimination, and device-resident boundary coordination. Across PEGASE-2869, PEGASE-9241, and ACTIVSg10k, the proposed multi-area GPU solver consistently achieved the best steady-state runtime among baselines while matching the converged solution quality.  In particular, it reached 24.7~ms on PEGASE-2869, 49.2~ms on PEGASE-9241, and 42.5~ms on ACTIVSg10k, corresponding to speedups of $3.47\times$, $6.90\times$, and $10.92\times$ over the centralized CPU (1T) baselines, and $3.35\times$, $5.85\times$, and $5.92\times$ over the aligned multi-area CPU baseline. These results indicate that SIMD-native, fully device-resident multi-area SE can provide consistent acceleration at realistic transmission-network scales.

The additional studies further clarify the method's operating regime. The partition-count sweep shows a moderate-$K$ sweet spot: partitioning improves throughput up to a point, but over-partitioning enlarges the boundary system and increases coordination cost. Taken together, these results indicate that high-performance GPU MASE depends not only on exposing parallelism, but also on aligning the solver backend with the sparse-interior, small-interface structure induced by the multi-area formulation.

Future work will focus on more interface-aware partitioning, more scalable boundary solvers for larger $K$, and lower per-iteration overhead through techniques such as CUDA graph capture. Another important direction is extending the framework to handle more irregular measurement dropouts and topology changes while preserving the efficiency of the fixed-template, device-resident implementation.

\begin{acks}
This material is based upon work supported by the
\grantsponsor{doe-oe}{U.S. Department of Energy's Office of Electricity}{https://www.energy.gov/oe/office-electricity}
under Award Number~\grantnum{doe-oe}{DE-OE0000985}. The views expressed herein do not necessarily represent the views of the U.S. Department of Energy or the United States Government. This material is based upon work supported by the
\grantsponsor{nsf}{National Science Foundation}{https://www.nsf.gov/}
under Award No.~\grantnum{nsf}{2348289}.
\end{acks}

\bibliographystyle{ACM-Reference-Format}
\bibliography{sample-base}

@String{Computing = "Computing" }

@article{gomez2011taxonomy,
  title={A taxonomy of multi-area state estimation methods},
  author={G{\'o}mez-Exp{\'o}sito, Antonio and De La Villa Ja{\'e}n, Antonio and G{\'o}mez-Quiles, Catalina and Rousseaux, Patricia and Van Cutsem, Thierry},
  journal={Electric Power Systems Research},
  volume={81},
  number={4},
  pages={1060--1069},
  year={2011},
  publisher={Elsevier}
}

@article{yilmaz2023robust,
  title={A robust parallel distributed state estimation for large scale distribution systems},
  author={Yilmaz, Ugur Can and Abur, Ali},
  journal={IEEE Transactions on Power Systems},
  volume={39},
  number={2},
  pages={4437--4445},
  year={2023},
  publisher={IEEE}
}

@book{abur2004power,
  title={Power system state estimation: theory and implementation},
  author={Abur, Ali and Exposito, Antonio Gomez},
  year={2004},
  publisher={CRC Press},
  address={Boca Raton, FL, USA}
}

@article{zhao2005multi,
  title={Multi area state estimation using synchronized phasor measurements},
  author={Zhao, Liang and Abur, Ali},
  journal={IEEE Transactions on Power Systems},
  volume={20},
  number={2},
  pages={611--617},
  year={2005},
  publisher={IEEE}
}

@article{guo2016hierarchical,
  title={Hierarchical multi-area state estimation via sensitivity function exchanges},
  author={Guo, Ye and Tong, Lang and Wu, Wenchuan and Sun, Hongbin and Zhang, Boming},
  journal={IEEE Transactions on Power Systems},
  volume={32},
  number={1},
  pages={442--453},
  year={2016},
  publisher={IEEE}
}

@article{shin2024accelerating,
  title={Accelerating optimal power flow with GPUs: SIMD abstraction of nonlinear programs and condensed-space interior-point methods},
  author={Shin, Sungho and Anitescu, Mihai and Pacaud, Fran{\c{c}}ois},
  journal={Electric Power Systems Research},
  volume={236},
  pages={110651},
  year={2024},
  publisher={Elsevier}
}

@article{lu2025cupdlp,
  title={cuPDLP. jl: A GPU implementation of restarted primal-dual hybrid gradient for linear programming in Julia},
  author={Lu, Haihao and Yang, Jinwen},
  journal={Operations Research},
  volume={73},
  number={6},
  pages={3440--3452},
  year={2025},
  publisher={INFORMS}
}

@article{wang2020privacy,
  title={Privacy-preserving hierarchical state estimation in untrustworthy cloud environments},
  author={Wang, Jingyu and Shi, Dongyuan and Chen, Jinfu and Liu, Chen-Ching},
  journal={IEEE Transactions on Smart Grid},
  volume={12},
  number={2},
  pages={1541--1551},
  year={2020},
  publisher={IEEE}
}

@article{fliscounakis2013contingency,
  title={Contingency ranking with respect to overloads in very large power systems taking into account uncertainty, preventive, and corrective actions},
  author={Fliscounakis, St{\'e}phane and Panciatici, Patrick and Capitanescu, Florin and Wehenkel, Louis},
  journal={IEEE Transactions on Power Systems},
  volume={28},
  number={4},
  pages={4909--4917},
  year={2013},
  publisher={IEEE}
}

@article{josz2016ac,
  title={AC power flow data in MATPOWER and QCQP format: iTesla, RTE snapshots, and PEGASE},
  author={Josz, C{\'e}dric and Fliscounakis, St{\'e}phane and Maeght, Jean and Panciatici, Patrick},
  journal={arXiv preprint arXiv:1603.01533},
  year={2016}
}

@misc{cudss2024,
  author = {{NVIDIA Corporation}},
  title  = {{NVIDIA cuDSS}: A High-Performance {CUDA} Library for Direct Sparse Solvers},
  year   = {2026},
  url    = {https://docs.nvidia.com/cuda/cudss/}
}

@misc{multi_area_se_repo,
  author       = {{Yifei Xu}},
  title        = {{Multi-Area State Estimation Testbed}},
  year         = {2026},
  howpublished = {\url{https://github.com/yifeihsu/multi_area_se}},
  note         = {GitHub repository, accessed April 4, 2026}
}

@misc{cuda_toolkit,
  author = {{NVIDIA Corporation}},
  title  = {{NVIDIA CUDA Toolkit}},
  year   = {2026},
  version = {13.2},
  url    = {https://developer.nvidia.com/cuda-toolkit}
}

@article{birchfield2016grid,
  title={Grid structural characteristics as validation criteria for synthetic networks},
  author={Birchfield, Adam B and Xu, Ti and Gegner, Kathleen M and Shetye, Komal S and Overbye, Thomas J},
  journal={IEEE Transactions on power systems},
  volume={32},
  number={4},
  pages={3258--3265},
  year={2016},
  publisher={IEEE}
}

@ArtifactSoftware{R,
    title = {R: A Language and Environment for Statistical Computing},
    author = {{R Core Team}},
    organization = {R Foundation for Statistical Computing},
    address = {Vienna, Austria},
    year = {2019},
    url = {https://www.R-project.org/},
}
\end{document}